\documentclass{iau} 
\usepackage{graphicx}
\usepackage{url}

\title[Finding evolved stars with \textit{Gaia}] %% give here short title %%
{Finding evolved stars in the inner Galactic disk with \textit{Gaia}}

\author[Quiroga-Nu\~{n}ez et al.]   %% give here short author list %%
{L.~H.~Quiroga-Nu\~{n}ez$^{1,2}$
    H.~J.~van Langevelde$^{2,1}$
    Y.~M.~Pihlstr\"{o}m$^{3}$
     L.~O.~Sjouwerman$^{4}$
 \and A.~G.~A.~Brown$^{1}$}

\affiliation{$^1$ Leiden Observatory, Leiden University, \\ P.O. Box 9513, 2300 RA Leiden, The Netherlands. \\ emails: {\tt quiroganunez@strw.leidenuniv.nl}; {\tt brown@strw.leidenuniv.nl} \\[\affilskip]
$^2$Joint Institute for VLBI ERIC (JIVE), \\ Postbus 2, 7990 AA Dwingeloo, The Netherlands. \\email: {\tt langevelde@jive.eu} \\[\affilskip]
$^3$Department of Physics and Astronomy, University of New Mexico, \\ MSC07 4220, Albuquerque, NM 87131, USA. \\email: {\tt ylva@unm.edu} \\[\affilskip]
$^4$National Radio Astronomy Observatory, \\ P.O. Box 0, Lopezville Road 1001, Socorro, NM 87801, USA. \\email: {\tt lsjouwer@nrao.edu}
}

\pubyear{2017}
\volume{330}  %% insert here IAU Symposium No.
\jname{Astrometry and Astrophysics in the \textit{Gaia} sky}
\editors{A. Recio-Blanco, P. de Laverny, A. Brown \& T. Prusti.}
\begin{document}

\maketitle

\begin{abstract}
The Bulge Asymmetries and Dynamical Evolution (BAaDE) survey will provide positions and line-of-sight velocities of $\sim 20,000$ evolved, maser bearing stars in the Galactic plane. Although this Galactic region is affected by optical extinction, BAaDE targets may have \textit{Gaia} cross-matches, eventually providing additional stellar information. In an initial attempt to cross-match BAaDE targets with \textit{Gaia}, we have found more than 5,000 candidates. Of these, we may expect half to show SiO emission, which will allow us to obtain velocity information. The cross-match is being refined to avoid false positives using different criteria based on distance analysis, flux variability, and color assessment in the mid- and near-IR. Once the cross-matches can be confirmed, we will have a unique sample to characterize the stellar population of evolved stars in the Galactic bulge, which can be considered fossils of the  Milky Way formation.
\end{abstract}

\firstsection % if your document starts with a section,

\section{Motivation}

The characterization of the stellar population of the Galactic bulge represents a key piece to understand the morphology and dynamical evolution of the inner Galaxy. This stellar population is dynamically affected by a massive bar (e.g. \cite{Dwek95}) and recent studies have shown an X-shaped structure (e.g. \cite{Wegg13}), similar to what it is seen in extragalactic edge-on boxy bulges. Optical surveys ---notably \textit{Gaia}--- are limited due to optical extinction, and are not able to make unhindered stellar astrometric measurements in the Galactic bulge, which complicates the characterization of this stellar population.

Radio campaigns are not affected by extinction and can therefore provide complementary information to optical surveys, especially at low latitudes. The Bulge Asymmetries and Dynamical Evolution (BAaDE) project surveys red giant stars for SiO maser emission at 43 and 86 GHz with the VLA and ALMA, eventually providing positions and radial velocities of approximately 20,000 targets along the Galactic plane (\cite{Sjouwerman16}). The BAaDE survey aims to significantly improve the dynamical models using radio sources in regions not reachable with optical surveys. The BAaDE survey is expanding  the currently known stellar tracers in the inner Galaxy by a large number. 

Tests for dynamical models of the Galaxy require large samples of stars with accurate positions and velocities. More details can be derived if distances are tied to stellar velocities. Therefore, we present an initial attempt to cross-match BAaDE targets with \textit{Gaia} DR1, resulting in more than 5,000 matches. However, since BAaDE targets were selected based on mid-IR colors measured with the MSX mission (\cite{Sjouwerman09}, (2016)) ---where the positional uncertainty is up to 2 arcsec--- the cross-matched sample could be contaminated by false positives. After confirming the matches, we will have a sample with optical, IR and radio information that can be used to characterize the stellar populations in the inner Galaxy, as well as to test dynamical models. In particular, we can obtain the positions, proper motions, parallaxes, colors and periods from \textit{Gaia} DR2 (April 2018). Until then, we can use the \textit{Gaia} DR1 positions for cross-matching.

\begin{figure}[b]
% \vspace*{-2.0 cm}
\begin{center}
\resizebox{\hsize}{!}{\includegraphics{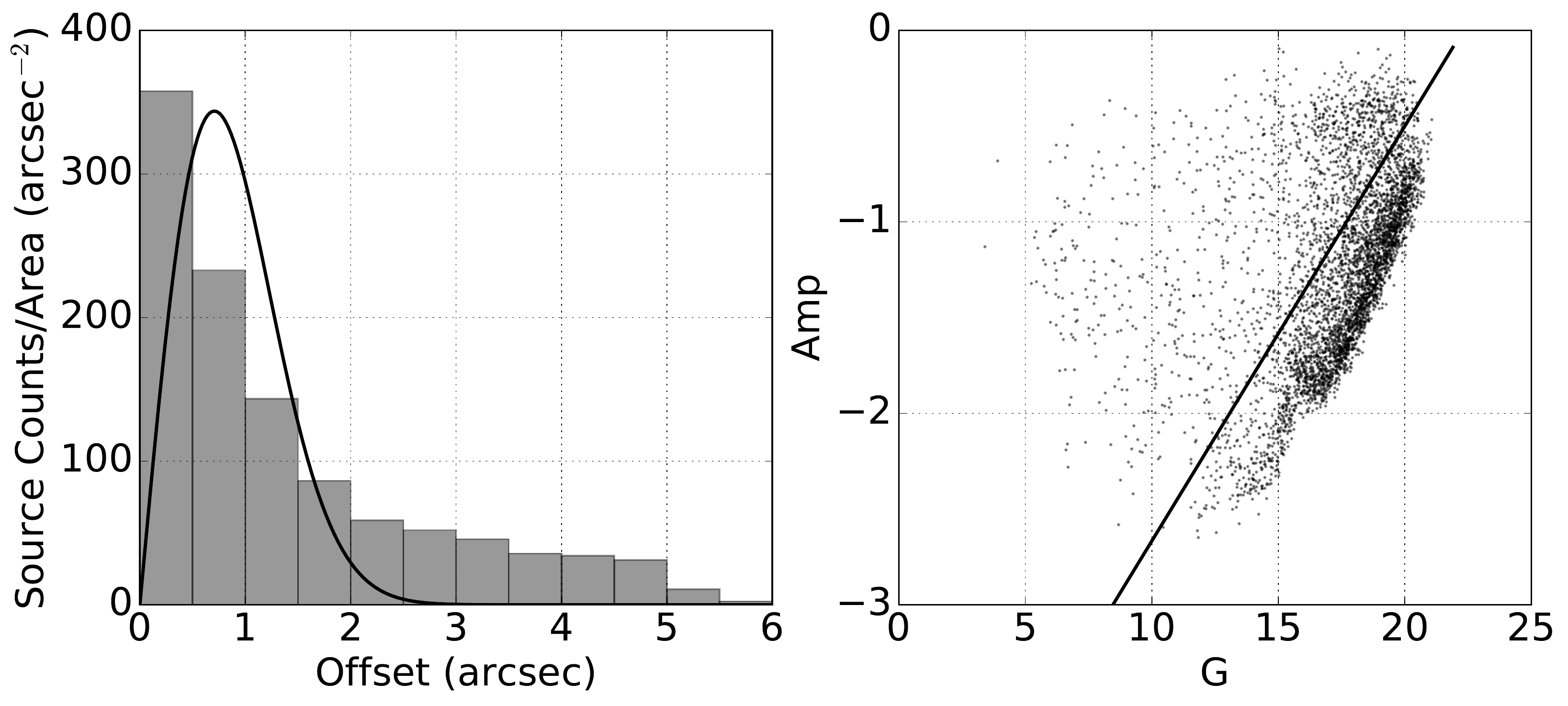}}
% \vspace*{-1.0 cm}
 \caption{\textbf{Left}: Offset distance between BAaDE and \textit{Gaia} sources. The solid line shows the offset distribution for 2 arcsec source position uncertainty, implying that sources with larger offsets may be false positives. \textbf{Right}: Amplitude-magnitude diagram for the cross-matches obtained between BAaDE and \textit{Gaia}. Higher amplitudes can be associate with pulsating AGB stars.}
   \label{fig1}
\end{center}
\end{figure}

\section{Cross-matching description}
\label{xmatch}

The BAaDE target selection was based on MSX colors, which in turn were based on IRAS color-color diagrams (see \cite{Sjouwerman09}). \cite{Veen88} developed an IRAS color-color diagram to study dust/gas envelopes (DGE) of Asymptotic Giant Branch (AGB) stars. They found that DGE stars appear in a sequence in the IRAS color-color diagram, perhaps associated with an evolutionary track with an increasing mass-loss rate. In this color-color diagram, SiO maser stars are found within a specific color regime, allowing a stellar selection based on the IRAS colors.  Later on, \cite{Sjouwerman09} were able to transform parts of this IRAS color-color diagram onto colors in the mid-IR, using MSX data.  With the improved angular resolution provided by MSX, red giant stars (with envelopes likely to harbor SiO maser emission) can be efficiently selected in the Galactic plane. 

To positionally match the BAaDE targets with other surveys, we consider a circular area with 5 arcsec radius around the BAaDE targets, based on the MSX positional accuracy (2 arcsec). Although the cross-match can be done directly with \textit{Gaia}, we initially cross-match BAaDE targets and 2MASS, because of three different reasons. Firstly, we do not expect that a target displaying both mid-IR emission (MSX) and optical emission (\textit{Gaia}) would not have emission in the near-IR (2MASS). Hence, by initially cross-matching with 2MASS, we are already avoiding some false positives. Secondly, the cross-match between 2MASS and \textit{Gaia} was already made by Marrese et al. (in preparation) using the best neighbor algorithm, finding more than 90$\%$ coincidences. Finally, 2MASS contributes with useful near-IR information to characterize the stellar population.

The cross-match between BAaDE targets and 2MASS produced more than 90$\%$ coincidences within 5 arcsec. However, looking at these sources in the \textit{Gaia} catalog, out of 5,674 coincidences seem to have a counter part at optical wavelengths. From those, 4,814 sources have only one \textit{Gaia} match and 860 sources have 2 or more \textit{Gaia} matches within the search radius. We will focus on the 4,814 sources that have a unique \textit{Gaia} match.

\section{False positive filters}
\label{false}

To refine the cross-matching by avoid false positives, several filtering methods have been considered. Below, we outline the most successful methods that we have applied.

\begin{figure}[b]
 %\vspace*{-2.0 cm}
\begin{center}\
\resizebox{\hsize}{!}{\includegraphics{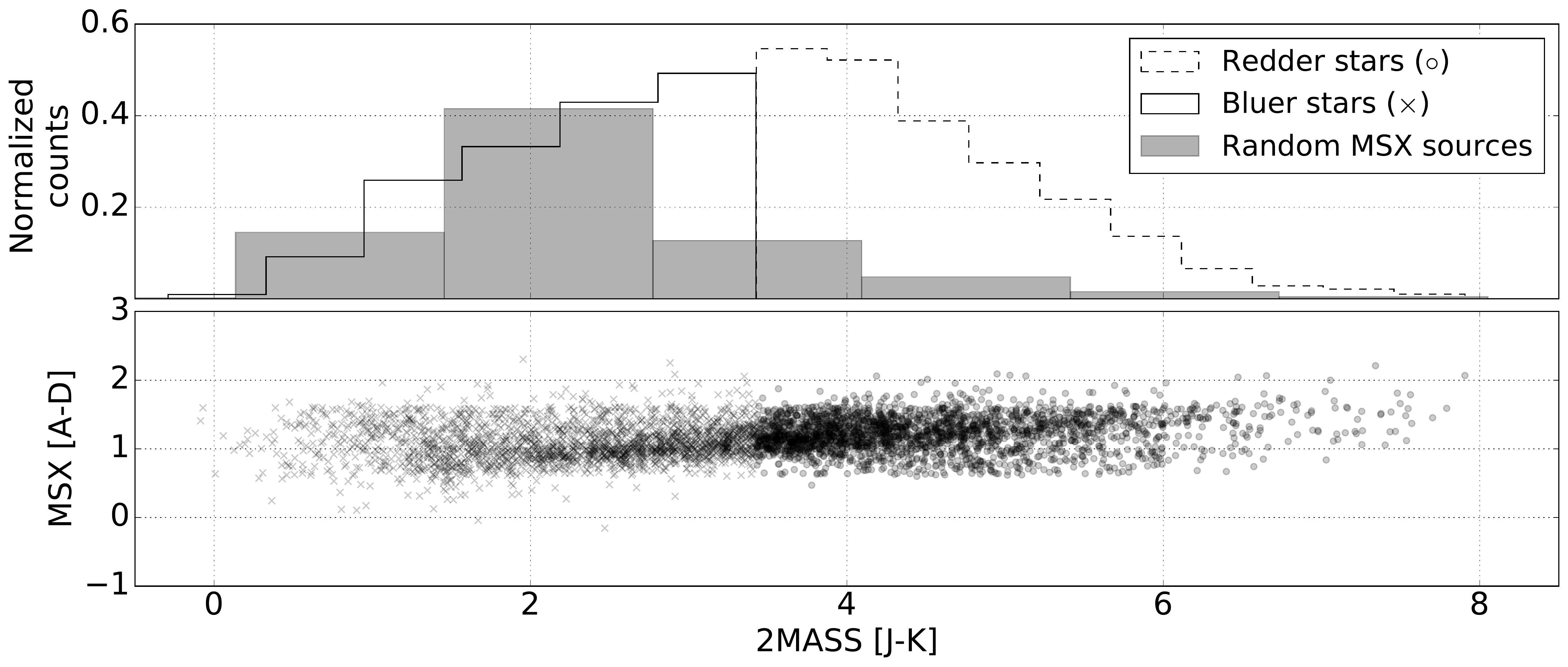}}
%\includegraphics[width=5in]{color_IAU.pdf} 
% \vspace*{-1.0 cm}
 \caption{\textbf{Lower}: Color-color diagram for  the cross-matches obtained between BAaDE targets and \textit{Gaia}. The sample was split in bluer (crosses) and redder sources (circles), see Sect.~\ref{color}. \textbf{Upper}: Histogram distributions for bluer and redder sources. The gray histogram represents a random sample of MSX sources showing that most of them are part of the bluer sample.}
   \label{fig2}
\end{center}
\end{figure}

\subsection{Distance analysis}
\label{dist}

The distribution of matches between the BAaDE targets and \textit{Gaia} shows Gaussian distributions for both components $(\Delta \alpha \times cos(\delta), \Delta \delta)$ with absolute mean values $<0.2$ arcsec. This is 2D Gaussian distribution can be converted to a function of the distance offset, which is a first-order Bessel function assuming the same standard deviations in both components. The left panel of Fig.~\ref{fig1} shows the expected distribution for a radius of 2 arcsec (representative as the typical MSX positional uncertainty) as a solid line. Excess sources at offsets above $\sim2$ arcsec may be considered false positives.

\subsection{Color filters in the mid- and near-IR}
\label{color}

Since the cross-match was made through 2MASS, the near-IR filters ($J,H,K$) and the mid-IR (MSX bands) can be used for color-color diagrams. The lower panel of Fig.~\ref{fig2} shows the color-color diagram between [A-D] MSX bands and [J-K] 2MASS filters for the matches between BAaDE targets and \textit{Gaia}. We calculated the mean value for the 2MASS colors and we split the sample in two different subsamples, i.e., $[J-K]<3.6$ (bluer stars) and $[J-K]>3.6$ (redder stars). AGB stars are expected to have redder colors (represented by a steeper slope in their SED), and therefore we expect that the redder stars are more likely to be correct cross-matches.

Moreover, the upper panel of Fig.~\ref{fig2} shows the histograms for the bluer and redder stars respectively, plotted on top of the distribution for random subset of MSX sources. The plot shows that most of the MSX sources are indeed bluer stars, in agreement with that redder stars (representing half of our sample) are more rare and could more easily be associated with pulsating AGB stars. 

\subsection{Variability of evolved stars}
\label{var}

The observed variability of the optical g-band can be quantified with an amplitude measure, defined as $\rm{Amp=log_{10}(\sqrt{N_{obs}}\frac{\sigma_g}{g})}$, where $N_{obs}$ is the number of observations. \cite{Belokurov17} calculated the amplitude for different stellar populations in the LMC and SMC, and localized Mira variables in the upper region of the amplitude-magnitude plot. This implies for a given range in $G$ these variable stars have a higher value of Amp than non-variable source of the same brightness. The right panel of Fig.~\ref{fig1} shows an amplitude-magnitude plot for the matches between BAaDE targets and \textit{Gaia}. The solid line represent the typical behavior for most of the \textit{Gaia} sources. Stars with amplitudes higher than -1 are highly related with pulsating stars and hence could be confirmed as properly matched. In contrast, Stars with amplitudes lower than -1 must be carefully reviewed by an alternative criterion for false positives.

\subsection{Statistical arguments}

Assuming an uniform distribution of sources in the bulge for the \textit{Gaia} detections and for the BAaDE targets, one could calculate the number of sources that randomly will match given the resolution of each survey. We estimate that the number of random matches should be less 1,200, which is low compared with our finding of 5,674 matches. Moreover, in the statistical calculation we have assumed that there is no optical extinction that could limit the number of \textit{Gaia} sources. Therefore, the actual number of random matches should be much lower than 1,200, confirming that our cross-match is not a consequence of random matches of unrelated sources.

%%\section{Using TGAS parallaxes}
%%\label{TGAS}

%The Tycho-Gaia solution (TGAS) can provide some parallax and therefore distance for 156 sources over the 6,000 coincidences. This sample is explained in Sect.~\ref{TGAS}.

%\begin{figure}[b]
%% \vspace*{-2.0 cm}
%\begin{center}\
%\resizebox{\hsize}{!}{\includegraphics{TGAS_IAU.pdf}}
%%\includegraphics[width=5in]{color_IAU.pdf} 
%% \vspace*{-1.0 cm}
% \caption{Path of pre-solar grains from their stellar sources to the laboratory.}
%   \label{fig3}
%\end{center}
%\end{figure}

%\section{VLBI astrometry with \textit{Gaia}}
%Both for connected element (VLALMA) and VLBI

\begin{acknowledgements} 
This work has made use of data from the European Space Agency
mission \textit{Gaia}, processed by the \textit{Gaia} Data Processing and Analysis Consortium (DPAC). Funding for the DPAC has been provided by national institutions, in particular the institutions participating in the \textit{Gaia} Multilateral Agreement. This material is based upon work supported by the National Science Foundation under Grant Number 1517970.
\end{acknowledgements}

\end{document}